\documentclass{ws-p10x7}

\begin{document}

\title{Next-to-leading Order SUSY-QCD Calculation of Associated 
Production of Gauginos and Gluinos}

\author{Edmond L. Berger and T. M. P. Tait}

\address{High Energy Physics Division, Argonne National Laboratory, 
Argonne, IL 60439,
USA\\E-mail: berger@anl.gov, tait@hep.anl.gov}

\author{M. Klasen}

\address{II.~Institut f\"ur Theoretische Physik, Universit\"at Hamburg, 
         D-22761 Hamburg, Germany\\E-mail: michael.klasen@desy.de} 

\twocolumn[\maketitle\abstract{
Results are presented of a next-to-leading order calculation in
perturbative QCD of the production of charginos and neutralinos in
association with gluinos at hadron colliders.  Predictions for cross 
sections are shown at the energies of the Fermilab Tevatron and CERN 
Large Hadron Collider for a typical supergravity (SUGRA) model of the 
sparticle mass spectrum and for a light gluino model.}]
  \vspace*{-10.3cm}
  \noindent hep-ph/0009256 \\
  ANL-HEP-CP-00-097 \\
  DESY 00-134
  \vspace*{6.84cm}
\section{Motivation}
The mass spectrum in typical supergravity and gauge-mediated models of 
supersymmetry (SUSY) breaking favors much lighter masses for 
gauginos than for squarks. Because the masses are smaller, there is greater 
phase space at the Tevatron and greater partonic luminosities 
for gaugino pair production, and for associated production of gauginos 
and gluinos, than for squark pair production.  In this contribution, we 
summarize our recent calculations at next-to-leading order (NLO) in 
perturbative quantum chromodynamics (QCD) of the total and 
differential cross sections for 
associated production of gauginos and gluinos at hadron 
colliders\cite{letter,prdtext}.  Associated production offers a chance to 
study the parameters of the soft SUSY-breaking Lagrangian.  Rates are 
controlled by the phases of the $\tilde \chi$ and $\tilde g$ masses 
and by mixing in the squark and gaugino sectors.  In addition to the 
potentially large cross section for associated production, the leptonic 
decay of the gaugino makes this process a good candidate for mass 
determination of the gluino and for discovery or exclusion of an 
intermediate-mass gluino.
\section{NLO SUSY-QCD Formalism}
Associated production of a gluino and a gaugino proceeds in leading order
(LO) through a quark-antiquark initial state and the
exchange of an intermediate squark in the $t$-channel or $u$-channel.
At NLO, loop corrections must be included.  In addition, there are 
2 to 3 parton processes initiated either by quark-antiquark scattering, 
with a gluon radiated into the final state, or by quark-gluon scattering
with a light quark radiated into the final state. For the quark-antiquark 
initial state, the loop diagrams involve the exchange of intermediate 
Standard Model or SUSY particles in self-energy, vertex, or box diagrams. 
Ultraviolet 
and infrared divergences appear at the upper and lower boundaries of 
integration over unobserved loop momenta.  They are regulated dimensionally 
and removed through renormalization or cancellation with corresponding 
divergences in the 2 to 3 parton (real emission) diagrams that have an 
additional gluon radiated into the final state.  In addition to soft 
divergences, real emission contributions have collinear divergences that 
are factored into the NLO parton densities.  The full treatment is 
presented in our long paper\cite{prdtext}.

\section{Tevatron\,\,and\,\,LHC\,\,Cross\,\,Sections} 
To obtain numerical predictions for hadronic cross sections, we choose 
an illustrative SUGRA model with parameters $m_0=100$ GeV, $A_0$=300 GeV, 
tan $\beta$ = 4, and sign $\mu$ = +.  Because the gluino, gaugino, and 
squark masses all increase with parameter $m_{1/2}$ (but are insensitive 
to $m_0$), we vary $m_{1/2}$ between 100 and 400 GeV.  The resulting masses 
for $\tilde{\chi}_{1...4}^0$ vary between 31...162, 63...317, 211...665, 
and 241...679 GeV; $\tilde{\chi}_{1,2}^\pm$ are almost degenerate in 
mass with $\tilde{\chi}^0_{2,4}$. The mass $m_{\tilde{\chi}_3^0} < 0$ 
inside a polarization sum.  Our approach is general, and results can be 
obtained for any set of gaugino and gluino masses.  For our second model, 
we select one\cite{raby} with an intermediate-mass gluino as the 
lightest SUSY particle (LSP), fixing 
$m_{\tilde{g}} =$ 30 GeV, and $m_{\tilde{q}} =$ 450 GeV.  We choose 
a weak sector identical to the SUGRA case.  In our paper\cite{prdtext}, we also 
quote results for anomaly mediated, gauge mediated, and gaugino 
mediated models.

\begin{figure}
\epsfxsize220pt
\figurebox{}{}{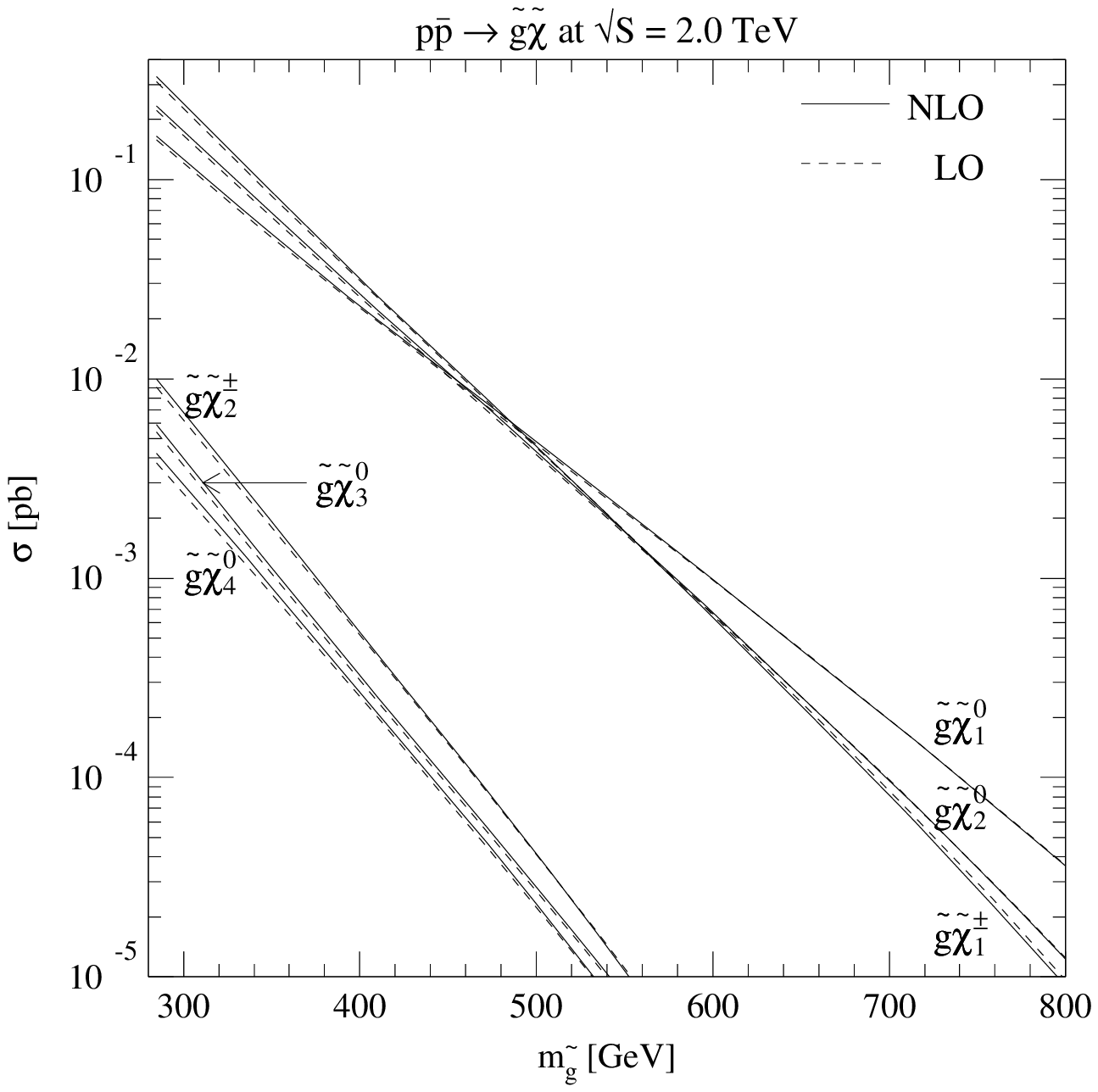}
\epsfxsize220pt
\vspace*{-1.0cm}
\figurebox{}{}{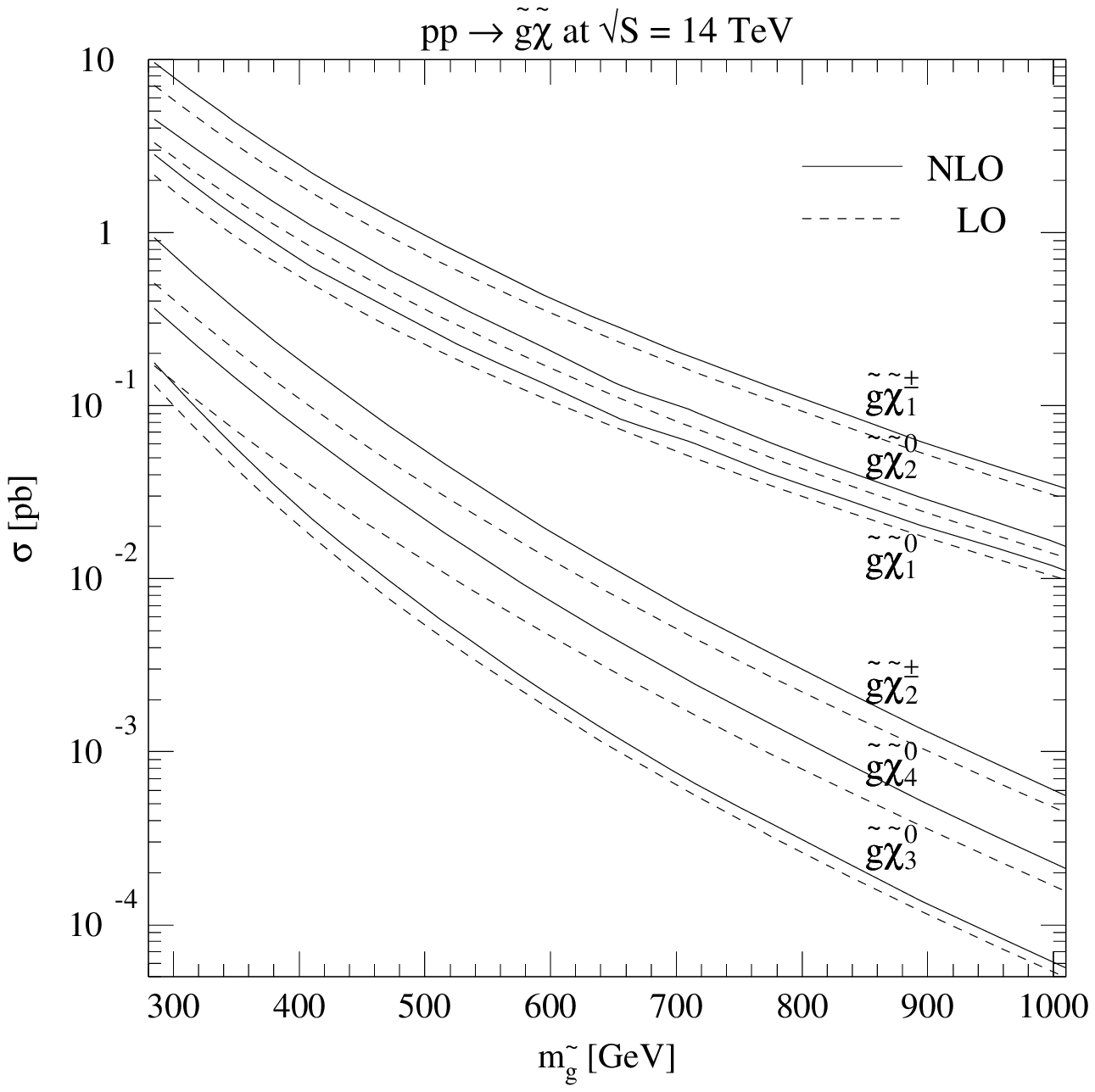}
\vspace*{-0.73cm}
\caption{Predicted total hadronic cross sections at Run II of the Tevatron 
and at the LHC for all six $\tilde{g} \tilde{\chi}$ channels in a typical SUGRA 
model as functions of the gluino mass.}
\label{fig:xsecsugra}
\end{figure}

We convolve LO and NLO partonic cross sections with CTEQ5 parton densities 
in LO and NLO ($\overline{\rm MS}$)
along with 1- and 2-loop expressions for $\alpha_s$, the corresponding
values of $\Lambda$, and five active quark flavors.  

For the SUGRA case, we present total hadronic cross sections in 
Fig.\ \ref{fig:xsecsugra} as functions of the gluino mass. The light 
gaugino channels should be observable at both colliders.  At the 
Tevatron, for $2~\rm{fb}^{-1}$ of integrated luminosity, 10 or more 
events could be produced in each of the lighter gaugino channels if 
$m_{\tilde g} < 450$ GeV.  The heavier Higgsino channels are suppressed 
by about one order of magnitude and might be observable only at the LHC.
As a rough estimate of uncertainty associated with the choice of 
parton densities, we note that the NLO cross section for 
$\tilde {\chi}_2^0$ production is lower by 12\% at the Tevatron 
with the CTEQ5 set than for the CTEQ4 set, and 4\% lower at the LHC.  
The impact of the NLO corrections can be seen more readily in the ratio 
of NLO to LO cross sections computed at a renormalizaton scale set equal 
to the average mass of the final state particles.  The NLO effects are 
moderate (of ${\cal O}$ (10\%)) at the Tevatron, while at the LHC the NLO 
contributions can increase the cross sections by as much as a factor of two. 
The second initial-state channel, initiated by gluon quark scattering, 
plays a significant role at the energy of the LHC. 

\begin{figure}
\epsfxsize220pt
\figurebox{120pt}{160pt}{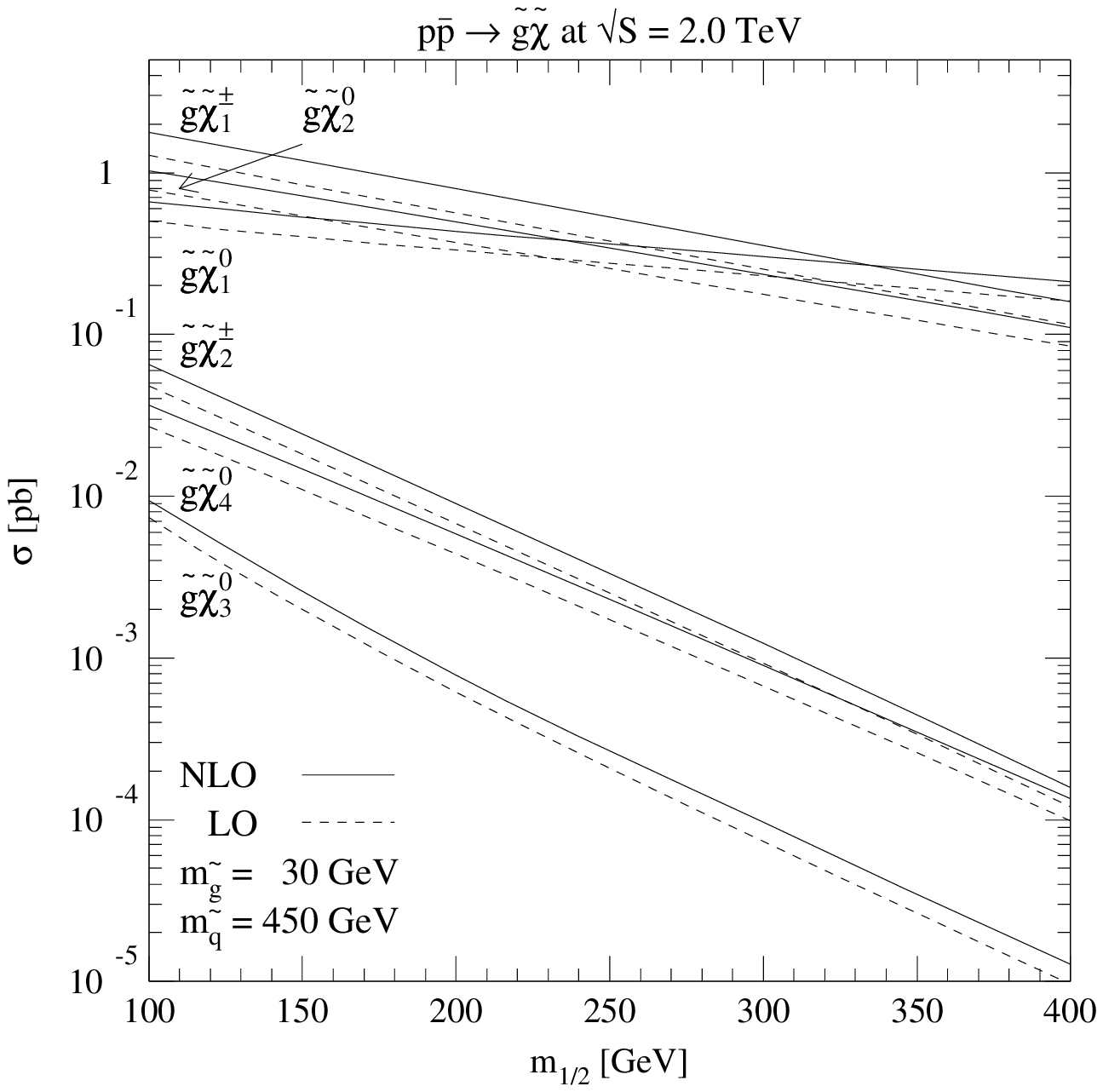}
\epsfxsize220pt
\vspace*{-1.0cm}
\figurebox{120pt}{160pt}{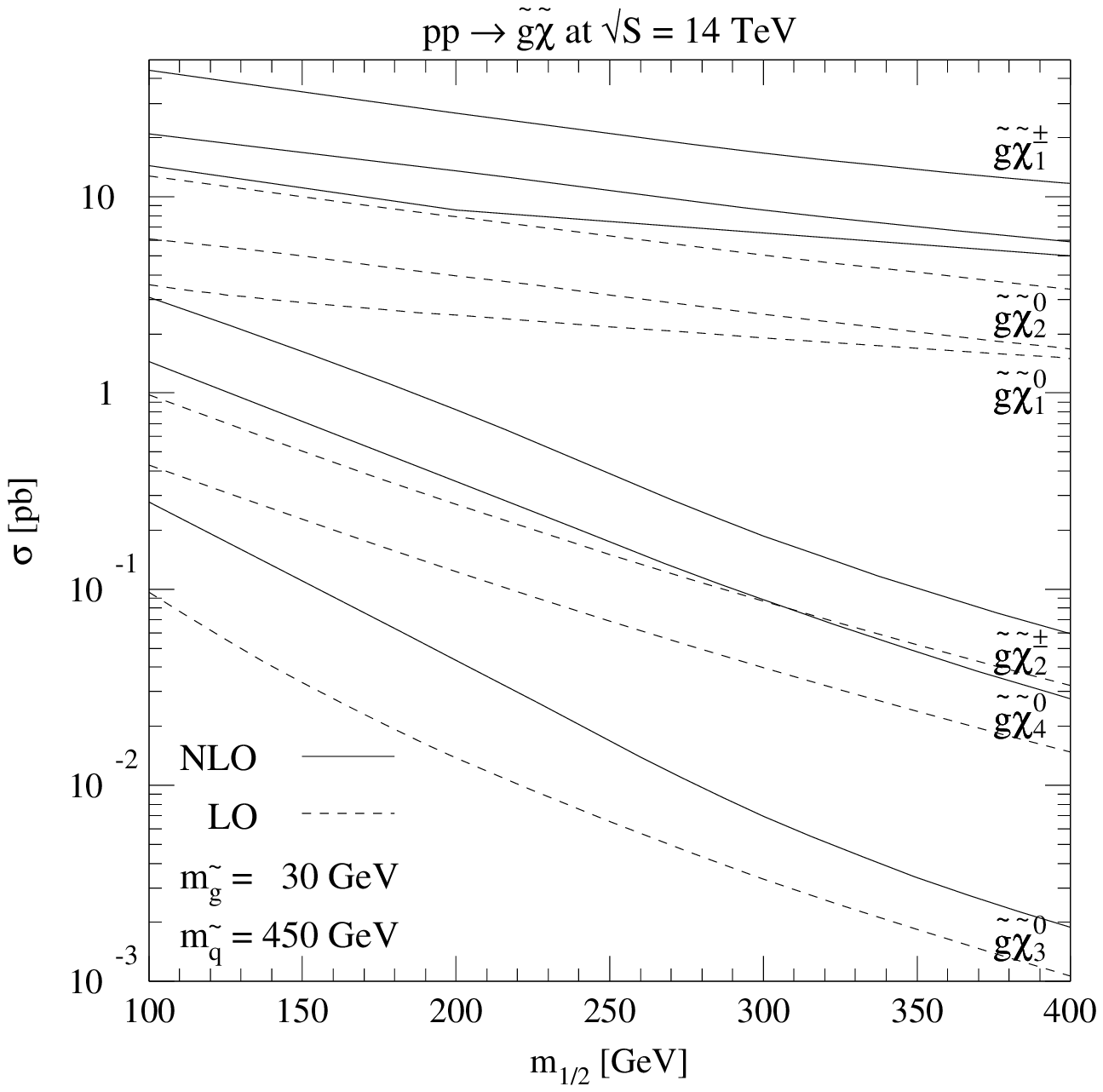}
\vspace*{-0.73cm}
\caption{Predicted total hadronic cross sections at Run II of the Tevatron 
and at the LHC for all six $\tilde{g} \tilde{\chi}$ channels in our model 
with a gluino of mass 30GeV, as functions of the parameter $m_{1/2}$.}
\label{fig:xseclite}
\end{figure}

For the case of a gluino with mass 30 GeV, the total hadronic cross 
sections are shown in Fig.\ \ref{fig:xseclite} as functions of $m_{1/2}$.  
At the Tevatron, for $2~\rm{fb}^{-1}$ of integrated luminosity, 100 or 
more events could be produced in each of the lighter gaugino channels if 
$m_{1/2} < 400$ GeV.  In this case, NLO enhancement factors lie in the 
ranges 1.3 to 1.4 at the Tevatron and 2 to 4 at the LHC.  

An important measure of theoretical reliability is the variation of 
the hadronic cross section with the renormalization and factorization 
scales.  At LO, these scales enter in the strong coupling constant
and the parton densities, while at NLO they appear also in the 
hard cross section. The scale dependence is reduced considerably after NLO 
effects are included.  The Tevatron (LHC) cross sections vary by 
$\pm 23 (12) \%$ at LO, but only by $\pm 8 (4.5) \%$ in NLO when the scale 
is varied by a factor of two around the central scale.

For experimental searches, distributions in transverse momentum are 
important since cuts on $p_T$ help to enhance the signal.  In our 
long paper\cite{prdtext}, we show that NLO contributions can have a large
impact on $p_T$ spectra at the LHC, owing to contributions from
the $gq$ initial state.  At the Tevatron the NLO $p_T$-distribution is 
shifted moderately to lower $p_T$ with respect to the LO
expectation.

\section*{Acknowledgments}
Work at Argonne National Laboratory is supported by the U.S. Department of 
Energy, Division of High Energy Physics, under Contract W-31-109-ENG-38. 
M.K. is supported by DFG through grant KL 1266/1-1.

\end{document}